# From One Electron to One Hole: Quasiparticle Counting in Graphene Quantum Dots Determined by Electrochemical and Plasma Etching

S. Neubeck[1], L. A. Ponomarenko[1], F. Freitag[1], A. J. M. Giesbers[2], U. Zeitler[2], S. V. Morozov[3], P. Blake[4], A. K. Geim[4], K. S. Novoselov[1,*]

[1]School of Physics & Astronomy, University of Manchester, Manchester, M13 9PL, UK
[2]High Field Magnet Laboratory, Institute for Molecules and Materials
Radboud University Nijmegen, Toernooiveld 7, 6525 ED Nijmegen, The Netherlands
[3]Institute for Microelectronics Technology, Chernogolovka, 142432, Russia
[4]Centre for Mesoscience and Nanotechnology, University of Manchester, Manchester, M13 9PL, UK

Graphene is considered to be a promising material for future electronics. The envisaged transistor applications often rely on precision cutting of graphene sheets with nanometer accuracy. In this letter we demonstrate graphene-based quantum dots created by using atomic force microscopy (AFM) with tip-assisted electrochemical etching. This lithography technique provides resolution of about 20 nm, which can probably be further improved by employing sharper tips and better humidity control. The behavior of our smallest dots in magnetic field has allowed us to identify the charge neutrality point and distinguish the states with one electron, no charge and one hole left inside the quantum dot.

[*] To whom correspondence should be addressed: Email: kostya@manchester.ac.uk

Recently obtained isolated graphene – a monolayer of carbon atoms packed into a hexagonal lattice – exhibits a range of extraordinary electronic properties (which originates from the linear, gapless spectrum of its quasiparticles[1-4]) and is widely considered as a promising material for future electronics.[1] At the same time, many electronics applications require the presence of an energy gap. To this end, considerable efforts were applied to create nanostructured devices out of graphene sheets (such as nanoribbons,[5,6] quantum point contacts (QPC),[7] single electron transistors,[1,8] quantum dots (QD)[7]), in which a gap can be opened due to quantum confinement of the charge carriers. In most cases, the formation of such graphene nanostructures relies on the removal of unwanted areas of graphene by reactive plasma etching (usually in oxygen plasma).[5-9] The performance of such nanostructured devices is expected to depend strongly on the quality[9,10] and the chemical nature of sample edges.[10] Therefore, it is crucially important to develop other methods of creating graphene nanostructures and control the edge orientation.

One of the possible alternatives for the reactive plasma etching is local electrochemical etching. Initially demonstrated for the case of graphite,[11-13] the local cutting by a biased conductive tip of an atomic force microscopy (AFM) system has been applied recently to graphene.[14] The technique is based on the dissociation of water molecules with subsequent chemical reaction of the radicals with the graphene carbon atoms. This opens up the possibility of local chemical modification of the graphene scaffolding, as well as for chemical modification of the edges. In this paper, we describe the fabrication of graphene QPCs and QDs by the AFM etching technique and measurements of their properties. This technique has allowed us to produce graphene structures with a resolution and quality similar to those previously achieved by using the high-resolution electron-beam lithography and subsequent plasma etching.[7] The operation of these devices is demonstrated by studying their behavior in magnetic field.

We used a Veeco Multimode scanning probe microscope with a NanoScope IIIa controller, which was operated in contact mode for AFM measurements and oxidation. The samples were mounted on a custom-made sample holder to allow for in-situ monitoring of their electronic properties. Topography scans revealed the height of our graphene samples (exfoliated from natural graphite[15]) to be typically about 0.8nm above the $SiO_2$ surface, which is the standard value for monolayer graphene in AFM measurements.[16] An air-tight enclosure was used during our experiments on local oxidation, which allowed us to maintain a constant temperature (22°C) and humidity (70%). Conductive silicon tips were biased with respect to the graphene samples (the bias was controlled by a Keithley source meter) and scanned along chosen lines with a speed of around 200nm $s^{-1}$ (higher speeds resulted in irregularly shaped etched structures or could even lead to the complete suppression of etching). We chose to work in the regime of "zero force" between the tip and the sample, which proved to produce the most reproducible results and thinnest cutting lines.

The direct current (DC) through the tip, as well as the resistance of the graphene devices (measured by the standard alternating-current lock-in technique), were monitored during the AFM oxidation. Typically, a tip bias of about -7V was required to initiate oxidation, which resulted in DC currents in the range of 10-100 nA. It should be noted that the threshold voltage was found to depend crucially on the humidity and biases in excess of -20V were required when the humidity dropped below 60%.

Figure 1 shows an example of the QD structure etched by this technique (the parameters used here were: bias -7V; humidity 70%; scanning speed 200nm s$^{-1}$). Bright (dark) lines on the top (middle) panel of Figure 1 are the nonconductive oxidized areas. Generally, applying a more negative bias to the tip for prolonged time (slower scanning) would result in complete etching away of graphene, rather than in oxidation. The central island of the QD (defined by the oxidized lines) is weakly connected by the narrow constrictions to the source and drain contacts. Using fresh tips and keeping the humidity relatively low, we managed to obtain oxidized lines with widths of down to 15nm, which is comparable to the best QPC and QD graphene structures obtained with electron-beam lithography.[7] The widths of the lines increased with increasing humidity (although the range of humidity where the technique works reliably is rather narrow at 60-80%) or if the bias voltage was significantly above its threshold value. The dependence of the line width on the applied AFM-tip bias voltage is shown in the bottom panel of Figure 1. Above the threshold (which depends on the humidity) the width of the etched lines increases approximately linearly with more negative bias voltages. This is in line with general expectations coming from a simple model of a water meniscus built up around the tip apex.[14]

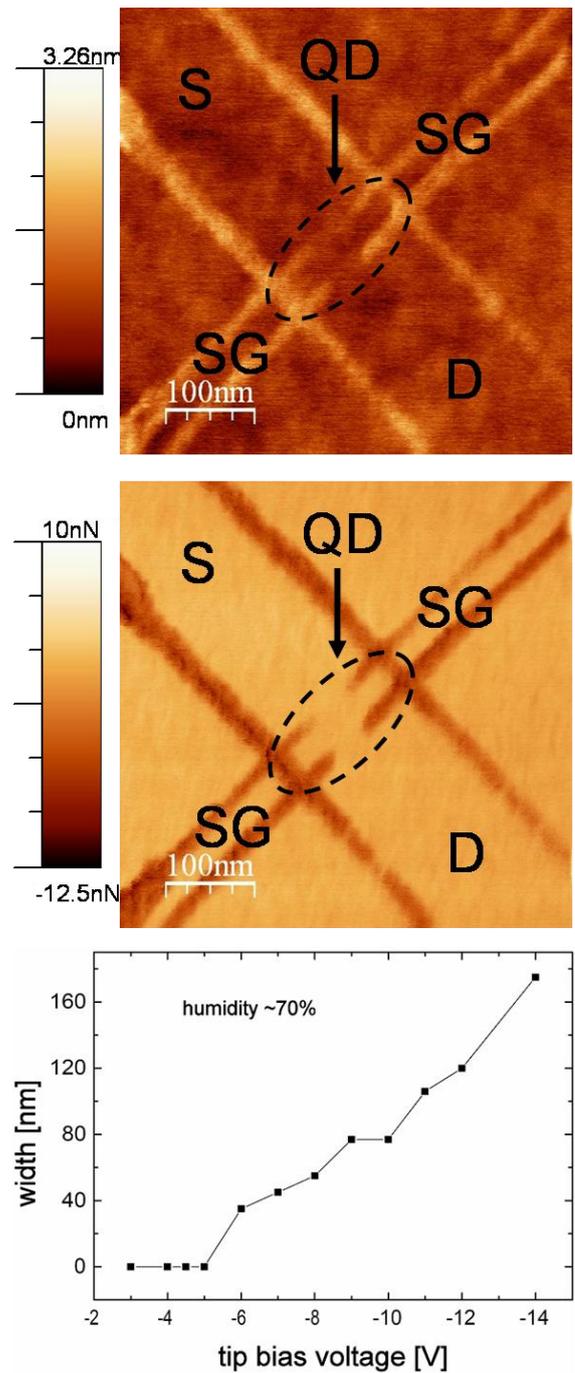

**Figure 1.** Top and middle: Example of a graphene QD structure created by local anodic oxidation. Top: Contact-AFM height image of a QD. Middle: The corresponding friction image. The bright regions in the friction image are intact graphene and the dark lines are the areas where graphene was etched away. The central island (marked as QD) is connected to the source (S) and drain (D) electrodes via narrow constrictions. Side gates (SG) are also formed from graphene. Bottom: The dependence of the width of the etched lines on the applied AFM-tip bias voltage.

An example of the typical behavior of the conductivity through one of our QDs (similar to the one shown on Figure 1) at various temperatures is presented in Figure 2. The conductivity shows a strongly distorted V-shape[7] behavior and its value drops well below one conductivity quantum ($e^2/h$) in the voltage range between 33 and 41V. In this range, several sharp conductivity peaks are observed. Outside this region, the conductivity grows above $0.5e^2/h$ (which we attribute to increased transparency of the constrictions (QPCs) between the QD and the source and drain contacts) and the QD levels could not be resolved. The nonmonotonic behavior of $G$ in the region below 33V and above 41V is probably due to changes in the transmission through the QPCs.[8]

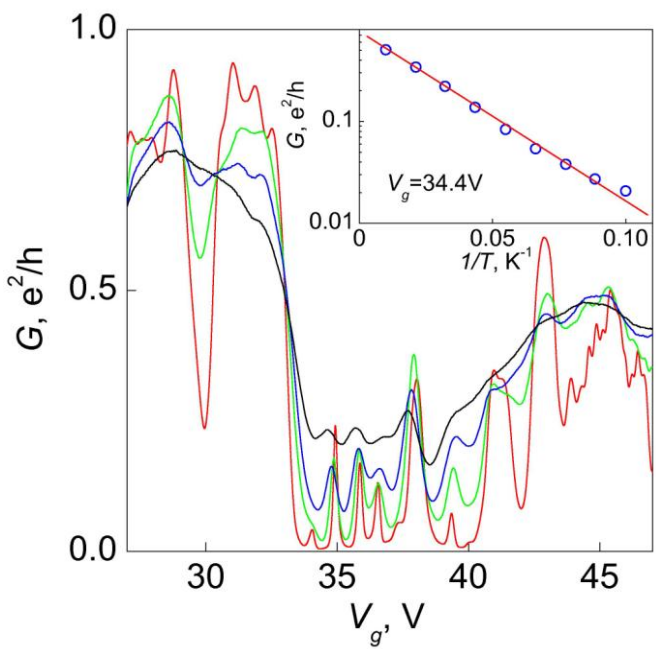

**Figure 2**. Conductivity through one of our QD devices (size of ≈100nm) as a function of backgate voltage for different temperatures. Red curve: 2.5K; green curve: 10K; blue curve: 20K; black curve: 30K. Inset: Temperature dependence of the conductivity at a minimum between peaks ($V_g$=34.4V).

The resonances in the voltage range between 33 and 41V are associated with the energy levels in the QD. The peaks are strongly aperiodic, which suggests that both the Coulomb energy and the size-quantization energies contribute to the splitting between the energy levels.[17] The number of peaks stays constant for temperatures below 20 K. Also, the relatively weak temperature dependence in the resonances (that can be as high as ~ 0.5 $e^2/h$) indicates that only one QD is present.[18] Peak height increases at lowest temperatures, as expected for a Coulomb blockade in the quantum case.[18] The temperature dependence of the minimum conductivity (Figure 2, inset) corresponds to an energy gap of 6.5meV, in agreement with the typical level spacing expected for a 100nm QD ($\delta E = \alpha/D$, where $\alpha$ varies around a value of 1 eV nm$^{-1}$ by a factor of 2 in different models[19,20]).

Similar values of the gap are obtained from the stability diagrams (conductivity versus the gate voltage and source-drain bias), such as the one presented in Figure 3, which shows the standard Coulomb diamonds.[21] The height of the diamonds directly yields the distance between adjacent energy levels and, for this particular sample, varies from 5 to 10mV. Such strong variation shows that the size quantization contributes significantly to the formation of energy levels in our small quantum dots.[17]

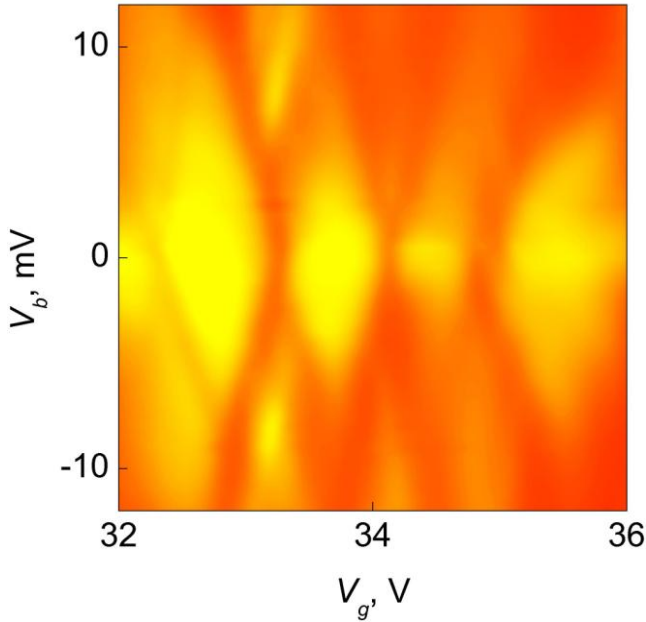

**Figure 3**. Coulomb diamonds for the QD in Figure 2 as a function of the gate voltage (the horizontal axis) and the source-drain bias voltage ($V_b$, vertical axis). $T$=2.5K. The conductivity varies from 0.002 $e^2/h$ (yellow) up to 0.2 $e^2/h$ (red).

In the remaining part of this paper, we demonstrate that our QDs can be set into the state with no charge (zero electrons and zero holes present). This also means that QDs (prepared by either AFM or electron-beam lithography) can be tuned into a state with any chosen number of electrons or holes. It has been proven previously[7,22] that graphene quantum dots allow for the observation of both electronic and hole states, although no method for determining of the exact crossover point has been demonstrated. In particular, the behavior of the energy levels in a magnetic field was used[22] in order to determine the approximate position of the electron/hole crossover in relatively large (80nm) QDs. Here, we will show that the same method, when applied to smaller QDs, gains enough resolution and allows for exact pin pointing of the crossover state with zero quasiparticles in a QD. The ability to set up a quantum dot in a state with a predetermined number of only a few electrons or holes might be extremely important for the realization of a particular spin state and thus for the implementation of qubits.[23]

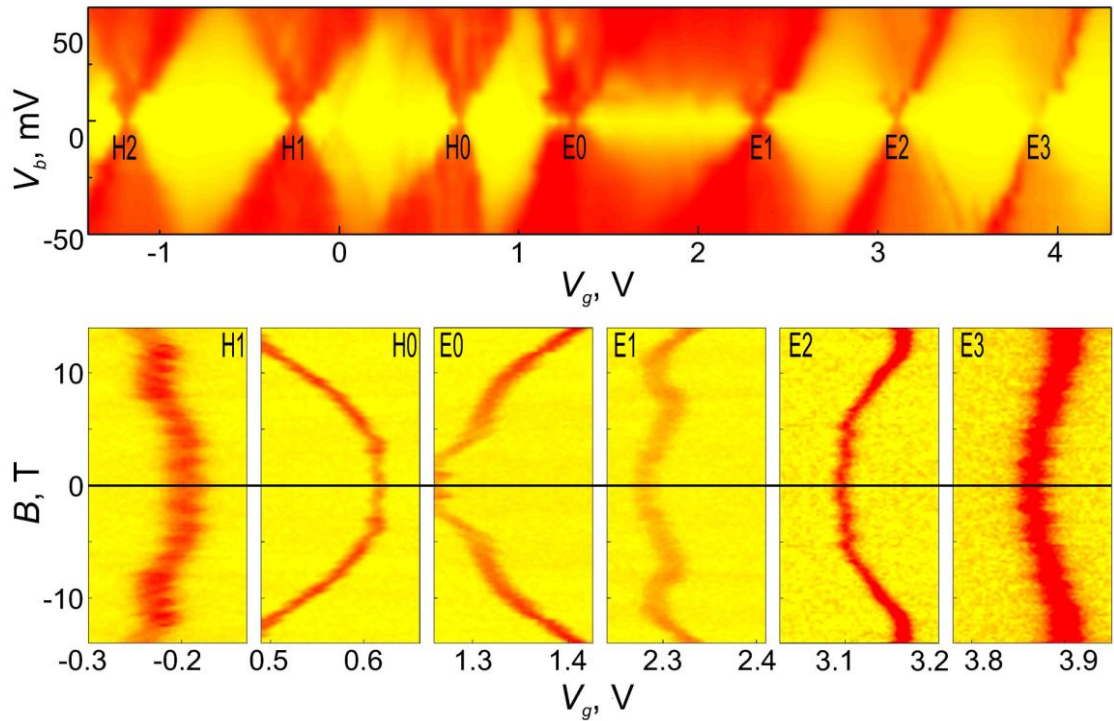

**Figure 4**. Top: Coulomb diamonds in another QD (≈20nm in size). $T=0.3K$; $B=0T$. Bottom: Evolution of the resonant-peak positions in magnetic field. $T=0.3K$, $V_b=0V$. The peaks are marked on the stability diagram in zero $B$ (top panel). The conductivity varies from practically zero (yellow) up to $0.05e^2/h$ (red).

The stability diagram for one of our devices is presented in Figure 4. Measurements at elevated temperatures indicate that the compensation point for this QD lies between -1V and 4V (for comparison, see Figure 2, in which the compensation point for that QD is expected between 34 and 39V). Accordingly, in our detailed experiments, we concentrated on the compensation region and followed the behavior of the conductivity peaks in a perpendicular magnetic field. The magnetic field causes a shift in the peak's positions as a function of gate voltage. Such behavior is presented in Figure 4. Importantly, there is a certain symmetry in the behavior of some peaks. For example, peak H0 shifts symmetrically with respect to E0, as with H1 to E1, H2 to E2, and so on. Furthermore, the behavior of E0 and H0 is notably different from all the others, including those that are not presented in Figure 4 (in total, we measured the behavior of about 20 peaks). Both E0 and H0 show monotonic and the largest shifts, whereas all the other peaks demonstrate weak nonmonotonic behavior.

Such magnetic response is well known for conventional quantum dots based on semiconducting heterostructures and can be explained in terms of the energy levels shifting in the magnetic field.[23] Depending on the orbital quantum number, up or down shifts in the energy position can be observed. The lowest level with zero orbital number always exhibits diamagnetic behavior (i.e., its energy increases with increasing magnetic field).[24] The nonmonotonic behavior of higher energy levels is normally explained in terms of crossing between levels in a magnetic field and many-body effects.[24] Following the same analysis, we attribute peaks E0 and H0 to the lowest levels for electrons and holes respectively. This implies that, in the range of gate voltages between 0.62 and 1.25V, the quantum dot is not charged, and for 0.25V ÷ 0.62V (1.25V ÷ 2.3V) the QD contains one hole (electron). The other conductivity peaks correspond to two (H1, E1), three (H2, E2), electrons or holes in the dot, and so on. The oscillatory behavior of the peak positions can be attributed to level crossing in the magnetic field.

It has been predicted[25] that, for quasiparticles with a Dirac-like spectrum in graphene, the first energy level for electrons (holes) should shift down (up) towards zero energy and eventually form zero Landau level in quantizing magnetic fields. However, this simple behavior is expected only in the absence of strong intervalley scattering so that the Dirac cones are preserved, and in sufficiently high fields such that $l_b << R$ (where $l_b = (\hbar/eB)^{1/2}$ is the magnetic length and $R$ the radius of a QD). The QDs presented in this work do not satisfy these criteria: the intervalley scattering is expected to be very efficient at the rough QD edges and the highest magnetic field we used ($B$=14T) yields $l_b$= 7 nm which is comparable with the QD's $R$=10nm. Lifting of the valley and spin degeneracies in graphene QDs and nanoribbons has also been experimentally demonstrated in previous experiments.[7,26] Under these conditions, it is reasonable to expect that the spatial quantization and strong intervalley scattering make quasiparticles "forget" about their initial Dirac-like spectrum and follow the standard behavior for massive electrons and holes. The diamagnetic shift of the first energy levels would then be described by the theory[24] developed for semiconducting quantum dots. At the moment, there is no theory developed for graphene with dominant intervalley scattering in order to compare our results.

In conclusion, the AFM lithography can be used to make graphene nanostructures with sizes below 15 nm. Furthermore, choosing the etching agent, one can control the functionalization of the edges of such structures. The resolution achieved by this technique is controlled by the humidity and the applied bias voltage. The behavior of our smallest quantum dots is strongly influenced by intervalley scattering at the sample edges, the regime that has not been discussed theoretically until recently.[27] Shifting of the peak position in a magnetic field allowed us to identify the empty state of the QD with no quasiparticles present and count the number of electrons and holes as the QD levels are filled in sequence. This is the first example of quantum dots in which a controllable ambipolar transition from a single hole, through an empty QD, to a state with a single electron has been achieved.

**Experimental Section**

Graphene crystallites were prepared from natural graphite[15] on an oxidized Si substrate (300 nm of $SiO_2$) by micromechanical cleavage.[1,2,16] Standard electron-beam lithography, thin-film deposition, and reactive plasma etching were then used to produce graphene Hall bars with a typical width of 1μm, having Ti/Au contacts.[2,3] Our samples were annealed for 4 h at $T=250^{o}C$ in a hydrogen/argon atmosphere (10% hydrogen) to remove resist residues. Electrical measurements revealed an electron mobility of ≈13,000 $cm^2$ $(V\ s)^{-1}$ (measured at typical carrier concentrations of $n=10^{12}$ $cm^{-2}$). The single-layer nature of the device used was confirmed by Raman spectroscopy and the observation of the half-integer quantum Hall Effect that is a characteristic signature for graphene.[3,4]

This work was supported by Engineering and Physical Sciences Research Council (UK), the Royal Society, the European Research Council (programs "Ideas", call: ERC-2007-StG and "New and Emerging Science and Technology," project "Structural Information of Biological Molecules at Atomic Resolution"), Office of Naval Research, and Air Force Office of Scientific Research. The authors are grateful to Nacional de Grafite for supplying high quality crystals of graphite.